\documentclass[aps,pra,reprint,showpacs,floatfix,longbibliography,superscriptaddress]{revtex4-1}
\usepackage{graphicx, amsmath, amssymb}
\usepackage[caption=false]{subfig}
\allowdisplaybreaks[1] 

\begin{document}


\newcommand{\braket}[2]{{\left\langle #1 \middle| #2 \right\rangle}}
\newcommand{\bra}[1]{{\left\langle #1 \right|}}
\newcommand{\ket}[1]{{\left| #1 \right\rangle}}
\newcommand{\ketbra}[2]{{\left| #1 \middle\rangle \middle \langle #2 \right|}}


\title{Quantum Walk Search on Kronecker Graphs}

\author{Thomas G.~Wong}
	\email{thomaswong@creighton.edu}
	\affiliation{Department of Physics, Creighton University, 2500 California Plaza, Omaha, NE 68178}

\author{Konstantin W{\"u}nscher}
	\email{konstantin.wuenscher.16@ucl.ac.uk}
	\affiliation{Department of Statistical Science, University College London, 1-19 Torrington Place, London, WC1E 7HB}
    
\author{Joshua Lockhart}
	\email{joshualockhart@gmail.com} 
	\affiliation{Department of Computer Science, University College London, Gower Street, London WC1E 6BT}

\author{Simone Severini}
	\email{s.severini@ucl.ac.uk}
	\affiliation{Department of Computer Science, University College London, Gower Street, London WC1E 6BT}
	\affiliation{Institute of Natural Sciences, Shanghai Jiao Tong University, Shanghai, China}

\begin{abstract}
	Kronecker graphs, obtained by repeatedly performing the Kronecker product of the adjacency matrix of an ``initiator'' graph with itself, have risen in popularity in network science due to their ability to generate complex networks with real-world properties. In this paper, we explore spatial search by continuous-time quantum walk on Kronecker graphs. Specifically, we give analytical proofs for quantum search on first-, second-, and third-order Kronecker graphs with the complete graph as the initiator, showing that search takes Grover's $O(\sqrt{N})$ time. Numerical simulations indicate that higher-order Kronecker graphs with the complete initiator also support optimal quantum search.
\end{abstract}


\maketitle


\section{Introduction}

Grover's algorithm \cite{Grover1996} is a foundational algorithm in quantum computing \cite{NielsenChuang2000} that searches an unordered database of $N$ items in $O(\sqrt{N})$ time, thus offering a quadratic speedup over classical search. However, Benioff \cite{Benioff2002} notes that the runtime can be slower when searching a spatial region, since the time taken for a quantum robot or cellular automata \cite{Meyer1996a,Meyer1996b} to traverse the physical database must be taken into consideration. Since then, much research has explored how quickly quantum computers search various spatial regions. See \cite{AA2005,CG2004,AKR2005} for some seminal papers, and \cite{Wong5,Wong7,Chakraborty2016} for some recent results. The structure of the physical database can be encoded as a combinatorial graph, the goal being to find a particular ``marked'' vertex using the least possible number of queries to an oracle encoding the graph structure. Often, the quantum search is performed using a quantum walk \cite{Kempe2003}, which respects the locality of the graph.

Most of the graphs on which quantum search has been analyzed have translational symmetry or some other structure that makes the behavior of the quantum algorithm amenable to rigorous proof. These constraints mean that the graphs considered are often very different from real-world networked data (henceforth, ``networks''), which are typically \emph{small-world} \cite{Milgram1967}, meaning the number of edges between any pair of nodes is small. The degree distributions of real-world networks are often scale-free, heavy-tailed, or follow power laws \cite{Barabasi1999}. The question of how quantum walk search performs on real-world networks has received much less attention.

To attack this question, one might consider investigating the behavior of quantum search on real-world networked datasets (see, for example, the Stanford Large Network Dataset Collection \cite{snapnets}). However, we are mostly interested in how the runtime of the search algorithm depends on the number of nodes $N$, and since these datasets are typically static with a fixed number of nodes, they are not suitable for this purpose. To determine the dependence on $N$, one would need to extrapolate the network to the future or rewind a network to the past, which is made difficult from the fact that randomly removing vertices destroys the degree distribution of the network \cite{Stumpf2005}.

One solution to this is to use Kronecker graphs \cite{Leskovec2005a,Leskovec2005b,Leskovec2007,Leskovec2010} to generate  synthetic networks that have some or all of the aforementioned real-world properties. Kronecker graphs can be ``grown'' iteratively, to include as many vertices as desired, and so they are suitable for our purposes. To produce a Kronecker graph, one begins with an ``initiator'' graph of $M$ vertices and its associated adjacency matrix $A$, where $A_{uv} = 1$ if vertices $u$ and $v$ are adjacent, and $0$ otherwise. The $j$th order Kronecker graph is the graph whose adjacency matrix is $A^{\otimes j}$, the Kronecker or tensor product of $A$ with itself $j$ times, i.e.,
\begin{equation}
	\label{eq:def}
	A^{\otimes j} = \underbrace{A \otimes A \otimes \dots \otimes A}_{j\text{ times}}.
\end{equation}
This defines a graph with $N = M^j$ vertices. Kronecker graphs can also be made stochastic, and they have been used to accurately model the arXiv citation graph, the internet at the level of autonomous systems, citations of U.S.~patents, the coauthor network, and the trust network of Epinions \cite{Leskovec2010}.

Besides their ability to be fitted to real-world datasets, Kronecker graphs have another advantage over other methods for generating real-world networks, such as the preferential attachment model \cite{Barabasi1999}. As real-world networks grow, the number of edges they sprout is typically more than the number of nodes. This means the network gets more dense over time, and the effective diameter tends to shrink. The preferential attachment model does not capture this, but Kronecker graphs do \cite{Leskovec2005a,Leskovec2005b}.

In this paper, we report our first steps toward understanding how quantum walks search Kronecker graphs by analyzing the case where the initiator is the complete graph, or all-to-all network. We denote the complete graph of $M$ vertices by $K_M$, and the $j$th order Kronecker graph generated by it by $K_M^{\otimes j} = K_M \otimes K_M \otimes \dots \otimes K_M$. This is an ideal Kronecker graph to begin with because, as we will see subsequently, it is amenable to rigorous analysis. We give proofs that the first-, second-, and third-order Kronecker graphs with complete initiator can be searched by a continuous-time quantum walk in $\pi\sqrt{N}/2$ time, which is the optimal $O(\sqrt{N})$ runtime of Grover's algorithm \cite{BHMT2000}. A continuous-time quantum walk searches \cite{CG2004} by starting in a uniform superposition $\ket{s}$ over all $N$ vertices:
\begin{equation}
	\label{eq:s}
	\ket{\psi(0)} = \ket{s} = \frac{1}{\sqrt{N}} \sum_{i=1}^N \ket{i},
\end{equation}
where $\ket{1}$, $\ket{2}$, \dots, $\ket{N}$ are the computational basis states that label the vertices. Then the system evolves by Schr\"odinger's equation $i\,d\psi/dt = H \psi$ (with $\hbar = 1$) with Hamiltonian
\begin{equation}
	\label{eq:H}
	H = -\gamma A^{\otimes j} - \ketbra{w}{w},
\end{equation}
where $\gamma$ is the jumping rate (amplitude per time) of the quantum walk, and $\ket{w}$ denotes the marked vertex we are looking for \cite{CG2004}. We end by showing results of numerical simulations that suggest higher-order Kronecker graphs, with the complete initiator, can also be quickly searched in the same $\pi\sqrt{N}/2$ time. For search on deterministic and stochastic Kronecker graphs with incomplete initiators that produce graphs with real-world properties, we will report our findings separately.


\section{First Order}

\begin{figure}
\begin{center}
	\includegraphics{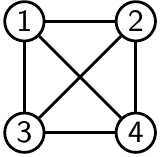}
	\caption{\label{fig:K4} $K_4$, the complete graph, or all-to-all network, of $4$ vertices.}
\end{center}
\end{figure}

In this section, we analyze search on the first-order Kronecker graph with complete initiator. From the definition of Kronecker graphs \eqref{eq:def} with $j = 1$, the first-order Kronecker graph is simply the complete initiator graph itself. For example, the complete graph with $4$ vertices is shown in Fig.~\ref{fig:K4}, and its adjacency matrix is
\[ A = \begin{pmatrix}
	0 & 1 & 1 & 1 \\
	1 & 0 & 1 & 1 \\
	1 & 1 & 0 & 1 \\
	1 & 1 & 1 & 0 \\
\end{pmatrix}. \]
If the initiator has $M$ vertices, then the first-order Kronecker graph has $N = M$ vertices.

\begin{figure}
\begin{center}
	\includegraphics{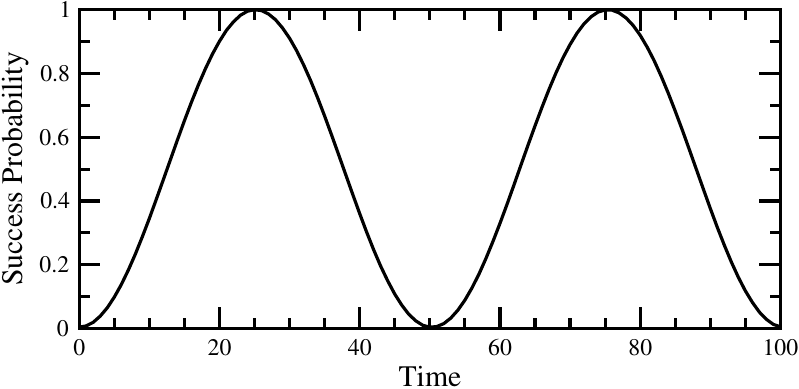}
	\caption{\label{fig:prob_time_j1} Success probability as a function of time for search by continuous-time quantum walk on the complete graph $K_M$ with $M = 256$.}
\end{center}
\end{figure}

Search on the complete graph is exactly the quantum walk formulation of Grover's unstructured search algorithm, since a complete graph constitutes an unstructured database. As shown in \cite{CG2004}, when the jumping rate $\gamma$ takes a ``critical value'' of $1/N$, the algorithm is equivalent to the ``analog analogue'' of Grover's algorithm \cite{FG1998}. Then the system evolves from the initial uniform superposition state $\ket{s}$ to the marked state $\ket{w}$ (up to a phase) in time $\pi\sqrt{N}/2 = O(\sqrt{N})$ \cite{Wong10}. Measuring the position of the walker at this time, one is certain to find it at the marked vertex. A simulation is shown in Fig.~\ref{fig:prob_time_j1}, and the success probability reaches $1$ at time $\pi\sqrt{256}/2 \approx 25.13$, as expected.

\section{Second Order}

Next, we consider the second-order Kronecker graph with complete initiator, which is also the line graph of a complete bipartite graph \cite{Godsil2017}. For example, if the initiator is the complete graph of $M = 4$ vertices from Fig.~\ref{fig:K4}, then the second order Kronecker graph has $N = M^2 = 4^2 = 16$ vertices, and its adjacency matrix is
\setcounter{MaxMatrixCols}{16}
\begin{align*}
	A^{\otimes 2} &= A \otimes A \\
	&= \begin{pmatrix}
		0 & A & A & A \\
		A & 0 & A & A \\
		A & A & 0 & A \\
		A & A & A & 0 \\
	\end{pmatrix} \\
	&= \begin{pmatrix}
		0 & 0 & 0 & 0 & 0 & 1 & 1 & 1 & 0 & 1 & 1 & 1 & 0 & 1 & 1 & 1 \\
		0 & 0 & 0 & 0 & 1 & 0 & 1 & 1 & 1 & 0 & 1 & 1 & 1 & 0 & 1 & 1 \\
		0 & 0 & 0 & 0 & 1 & 1 & 0 & 1 & 1 & 1 & 0 & 1 & 1 & 1 & 0 & 1 \\
		0 & 0 & 0 & 0 & 1 & 1 & 1 & 0 & 1 & 1 & 1 & 0 & 1 & 1 & 1 & 0 \\
		0 & 1 & 1 & 1 & 0 & 0 & 0 & 0 & 0 & 1 & 1 & 1 & 0 & 1 & 1 & 1 \\
		1 & 0 & 1 & 1 & 0 & 0 & 0 & 0 & 1 & 0 & 1 & 1 & 1 & 0 & 1 & 1 \\
		1 & 1 & 0 & 1 & 0 & 0 & 0 & 0 & 1 & 1 & 0 & 1 & 1 & 1 & 0 & 1 \\
		1 & 1 & 1 & 0 & 0 & 0 & 0 & 0 & 1 & 1 & 1 & 0 & 1 & 1 & 1 & 0 \\
		0 & 1 & 1 & 1 & 0 & 1 & 1 & 1 & 0 & 0 & 0 & 0 & 0 & 1 & 1 & 1 \\
		1 & 0 & 1 & 1 & 1 & 0 & 1 & 1 & 0 & 0 & 0 & 0 & 1 & 0 & 1 & 1 \\
		1 & 1 & 0 & 1 & 1 & 1 & 0 & 1 & 0 & 0 & 0 & 0 & 1 & 1 & 0 & 1 \\
		1 & 1 & 1 & 0 & 1 & 1 & 1 & 0 & 0 & 0 & 0 & 0 & 1 & 1 & 1 & 0 \\
		0 & 1 & 1 & 1 & 0 & 1 & 1 & 1 & 0 & 1 & 1 & 1 & 0 & 0 & 0 & 0 \\
		1 & 0 & 1 & 1 & 1 & 0 & 1 & 1 & 1 & 0 & 1 & 1 & 0 & 0 & 0 & 0 \\
		1 & 1 & 0 & 1 & 1 & 1 & 0 & 1 & 1 & 1 & 0 & 1 & 0 & 0 & 0 & 0 \\
		1 & 1 & 1 & 0 & 1 & 1 & 1 & 0 & 1 & 1 & 1 & 0 & 0 & 0 & 0 & 0 \\
	\end{pmatrix}.
\end{align*}

\begin{figure}
\begin{center}
	\subfloat[] {
		\includegraphics{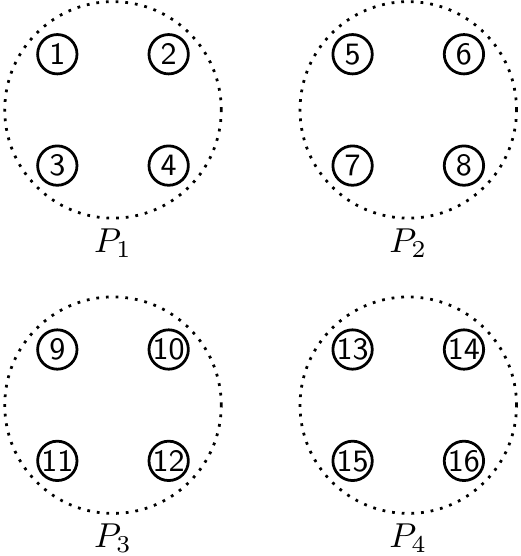}
		\label{fig:K4K4_inter}
	}

	\subfloat[] {
		\includegraphics{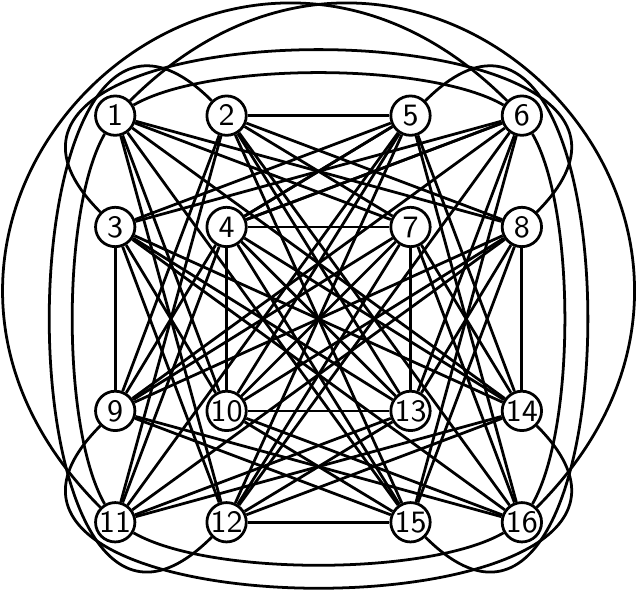}
		\label{fig:K4K4}
	}
	\caption{(a) Four sets of four vertices. (b) The second-order Kronecker graph $K_4 \otimes K_4$.}
\end{center}
\end{figure}

To draw this Kronecker graph, we begin with the initiator $K_4$ in Fig.~\ref{fig:K4} and replace each vertex with four new nodes, as shown in Fig.~\ref{fig:K4K4_inter}, grouping them in sets $P_1$ through $P_4$. Now, consider a specific vertex, say vertex $1$. First, it is not connected to any other vertices its set $P_1$. Next, note vertex $1$ is in the top-left corner of $P_1$, and similarly, vertex $5$ is in the top-left corner of $P_2$, vertex $9$ is in the top-left corner of $P_3$, and vertex $13$ is in the top-left corner of $P_4$. Vertex $1$ is adjacent to all vertices in the other partite sets except these that share the same top-left position. So vertex $1$ is adjacent to vertices $6$, $7$, and $8$ in $P_2$, vertices $10$, $11$, and $12$ in $P_3$, and vertices $14$, $15$, and $16$ in $P_4$. Following this procedure for each vertex, the Kronecker graph $K_4 \otimes K_4$ is shown in Fig.~\ref{fig:K4K4}. Note $P_1$, $P_2$, $P_3$, and $P_4$ are partite sets of a 4-partite graph.  

To determine how quickly a continuous-time quantum walk searches second-order Kronecker graphs with the complete initiator, we next prove that $K_M \otimes K_M$ is a strongly regular graph \cite{Cameron1991}. A strongly regular graph $(N,k,\lambda,\mu)$ is a graph of $N$ vertices where every vertex has $k$ neighbors, adjacent vertices share $\lambda$ common neighbors, and nonadjacent vertices share $\mu$ common neighbors. How quickly a continuous-time quantum walk searches a strongly regular graph, depending on its parameters, was investigated in \cite{Wong5}.

To determine the parameters of the strongly regular graph, let us work through the example of $K_4 \otimes K_4$ in Fig.~\ref{fig:K4K4}. Then, we will generalize it to arbitrary $K_M \otimes K_M$. First, every vertex is adjacent to $3$ vertices in $3$ partite sets, so the graph is regular with degree $3^2 = 9$. Second, adjacent vertices must be in different partite sets, and their positions (top-left, top-right, etc.) within their partite sets must also differ. For example, vertices $1$ and $6$ are adjacent. There are $4-2 = 2$ remaining partite sets in which they have mutual neighbors, and each of these partite sets has $4-2 = 2$ positions within them containing mutual neighbors. So adjacent vertices have $2^2 = 4$ mutual neighbors. Finally, nonadjacent vertices could be in the same partite set or in different partite sets. If the nonadjacent vertices are in the same partite set, such as vertices $1$ and $2$, then there are $4-1 = 3$ other partite sets that each contain $4-2 = 2$ mutual neighbors for a total of $3 \cdot 2 = 6$ mutual neighbors. If they are in different partite sets, such as vertices $1$ and $5$, then they must occupy the same position within their respective sets, such as the top-left corner. This leaves $4-2 = 2$ other partite sets that each contain $4-1=3$ mutual neighbors, for a total of $2 \cdot 3 = 6$ mutual neighbors. So the number of mutual neighbors of nonadjacent vertices is the same in both cases. Combining these results, $K_4 \otimes K_4$ is strongly regular with parameters are $(N,k,\lambda,\mu) = (16,9,4,6)$.

Generalizing this, $K_M \otimes K_M$ is an $M$-partite graph that is strongly regular with parameters $(N,k,\lambda,\mu)$, where
\begin{align*}
	N &= M^2, \\
	k &= (M-1)^2, \\
	\lambda &= (M-2)^2, \\
	\mu &= (M-1)(M-2).
\end{align*}

\begin{figure}
\begin{center}
	\includegraphics{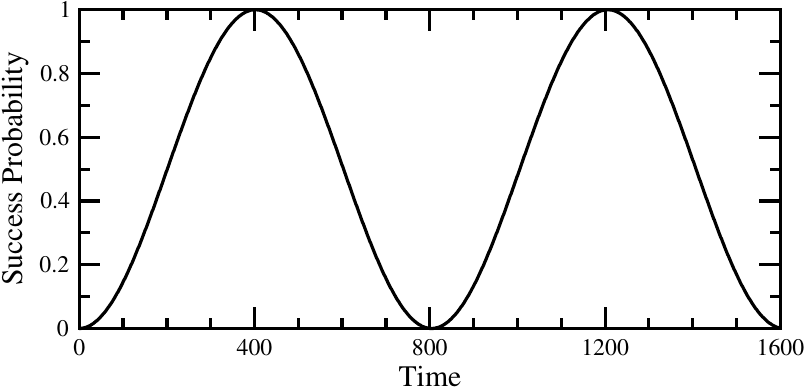}
	\caption{\label{fig:prob_time_j2} Success probability as a function of time for search by continuous-time quantum walk on $K_M \otimes K_M$ with $M = 256$.}
\end{center}
\end{figure}

From \cite{Wong5}, if the parameters of a strongly regular graph satisfy $k = o(N)$ and $k = o[(\mu N)^{2/3}]$, then when the jumping rate $\gamma$ takes a ``critical value'' of $1/k + 1/[(N-1)\mu]$, the continuous-time quantum walk searches the graph with probability $1$ at time $\pi \sqrt{N}/2$, asymptotically. With the parameters we derived for $K_M \otimes K_M$, $k$ scales as $\sqrt{N}$, and $(\mu N)^{2/3}$ scales as $N^{4/3}$. So both conditions that $k = o(N)$ and $k = o[(\mu N)^{2/3}]$ are satisfied, and a continuous-time quantum walk searches the second-order Kronecker graph with complete initiator in $\pi\sqrt{N}/2$ time. A simulation is shown in Fig.~\ref{fig:prob_time_j2}, and the success probability reaches $1$ at time $\pi\sqrt{256^2}/2 \approx 402.12$, as expected.

\section{Third Order}

\begin{figure*}
\begin{center}
	\includegraphics{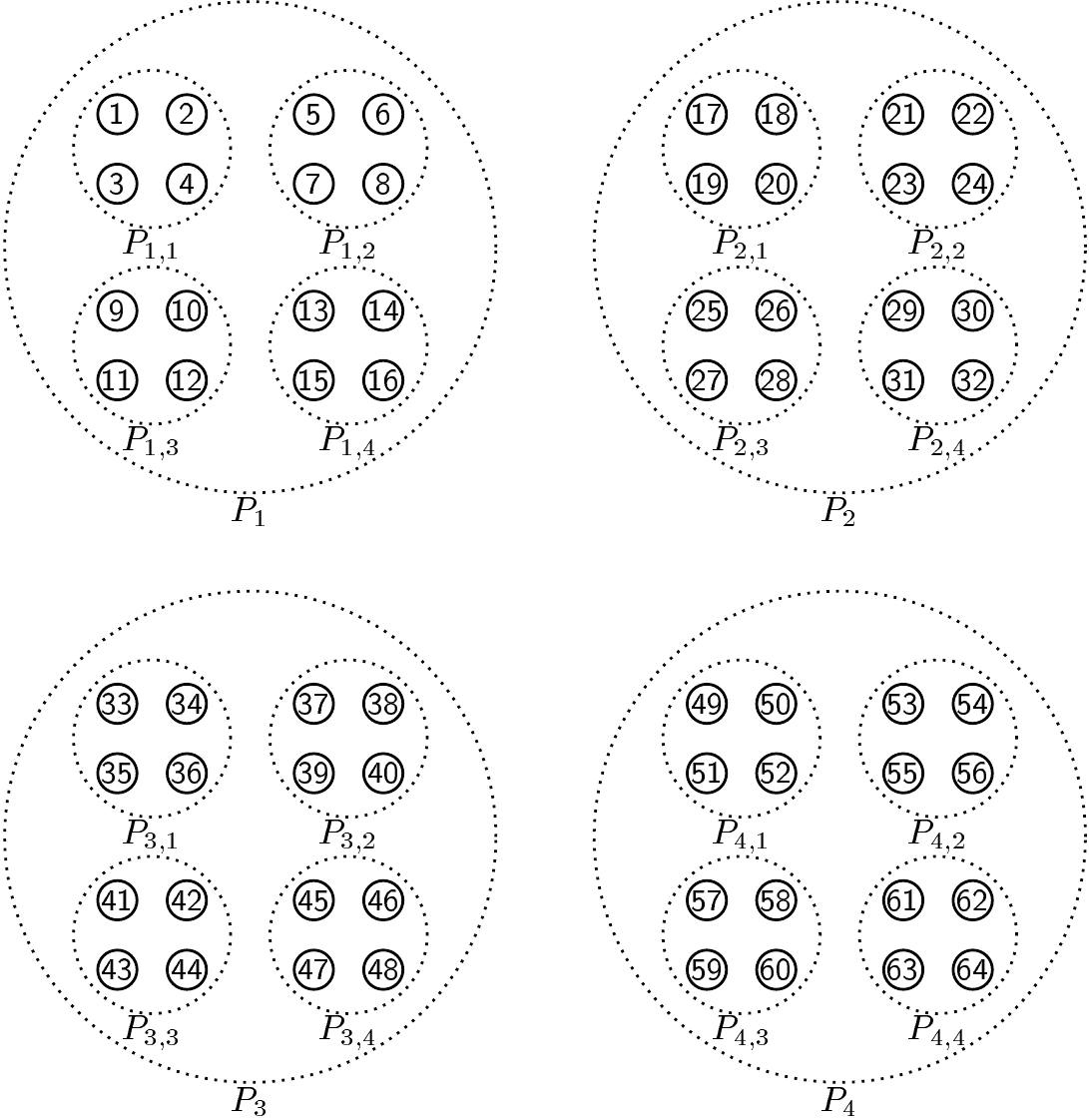}
	\caption{\label{fig:K4K4K4_inter} The vertices of $K_4 \otimes K_4 \otimes K_4$, arranged in four sets $P_i$ with $i = 1,2,3,4$, each with four subsets $P_{i,j}$ with $j = 1,2,3,4$.}
\end{center}
\end{figure*}

Now, we consider the third-order Kronecker graph. For example, using $K_4$ from Fig.~\ref{fig:K4} as the initiator, the third-order Kronecker graph has $N = M^3 = 4^3 = 64$ vertices, and its adjacency matrix is
\begin{align*}
	A^{\otimes 3} 
		&= A \otimes A \otimes A \\
		&= \begin{pmatrix}
			0 & A^{\otimes 2} & A^{\otimes 2} & A^{\otimes 2} \\
			A^{\otimes 2} & 0 & A^{\otimes 2} & A^{\otimes 2} \\
			A^{\otimes 2} & A^{\otimes 2} & 0 & A^{\otimes 2} \\
			A^{\otimes 2} & A^{\otimes 2} & A^{\otimes 2} & 0 \\
		\end{pmatrix}.
\end{align*}

To draw this Kronecker graph, we again begin with the initiator $K_4$ from Fig.~\ref{fig:K4} and replace each vertex with the sixteen vertices of $K_4 \otimes K_4$ from Fig.~\ref{fig:K4K4_inter}. The result of this substitution is shown in Fig.~\ref{fig:K4K4K4_inter}, without any edges. In this figure, we grouped together sets $P_i$ of sixteen vertices, and further grouped subsets $P_{i,j}$ of four vertices.

\begin{figure*}
\begin{center}
	\includegraphics{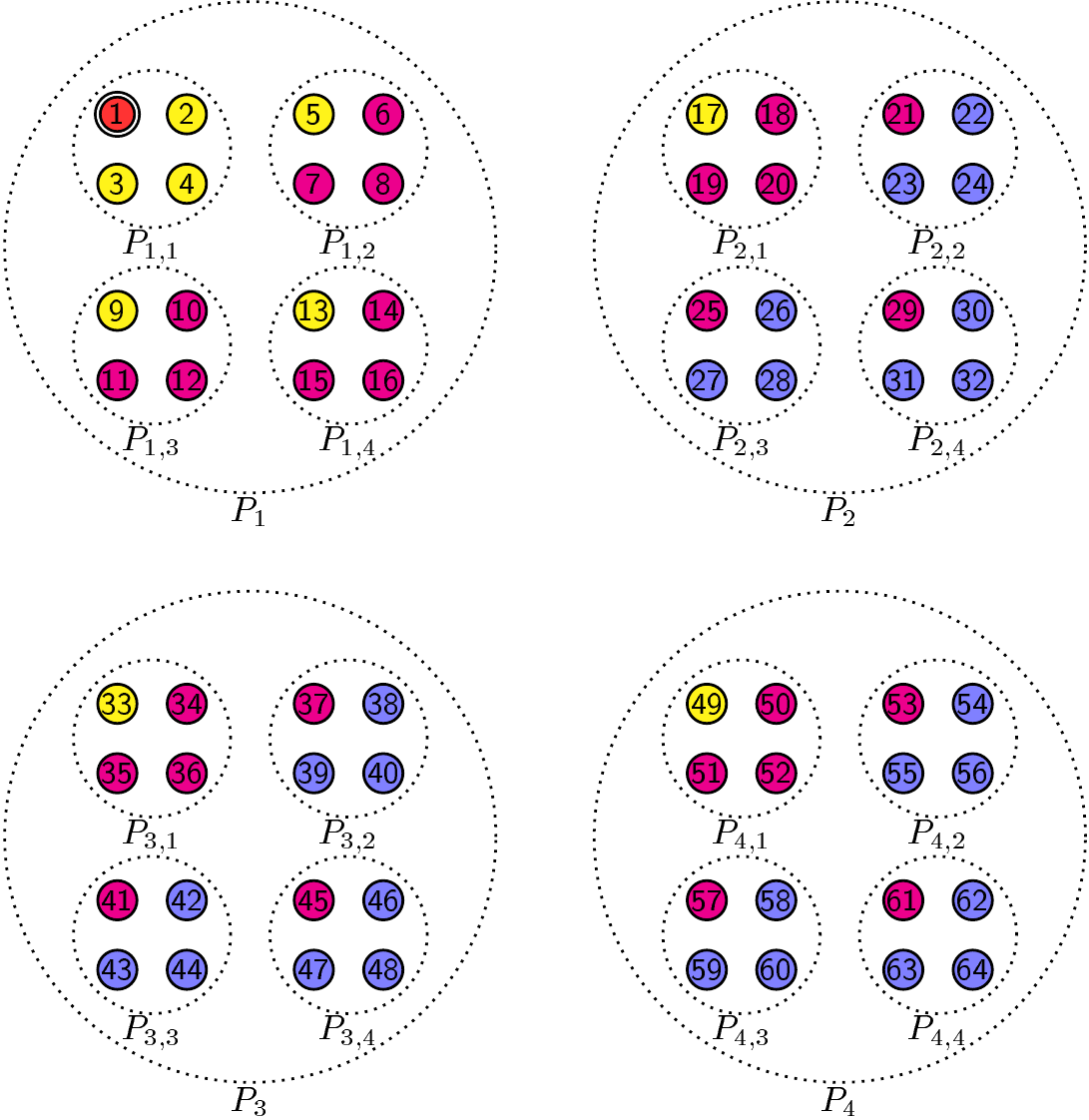}
	\caption{\label{fig:K4K4K4_v1_subspace} The vertices of $K_4 \otimes K_4 \otimes K_4$, arranged in four sets $P_i$ with $i = 1,2,3,4$, each with four subsets $P_{i,j}$ with $j = 1,2,3,4$. Vertex 1 is marked, colored red, and identified by a double circle. Its neighbors are colored blue, and its nonadjacent vertices are colored yellow and magenta, depending on their mutual neighbors. Identically colored vertices evolve identically.}
\end{center}
\end{figure*}

To determine the edges, consider a specific vertex, say vertex 1, which is in the top-left corner of its subset $P_{1,1}$. Then, vertex 1 is nonadjacent to any vertex in the top-left position of its subset, so it is nonadjacent to vertices 5, 9, 13, 17, 21, 25, 29, 33, 37, 41, 45, 49, 53, 57, and 61. Vertex 1 is in subset $P_{1,1}$, which is the top-left subset within the set $P_1$. Then, vertex 1 is nonadjacent to the subsets $P_{1,1}$, $P_{2,1}$, $P_{3,1}$, and $P_{4,1}$, since those are the top-left subsets of their partite sets. Finally, vertex $1$ is within set $P_1$, so it is not adjacent to any other vertex in $P_1$. Vertex 1 is adjacent to everything else. Explicitly drawing all the edges is messy, so we do not do so. In Fig.~\ref{fig:K4K4K4_v1_subspace}, however, the vertices adjacent to vertex $1$ are colored blue.

Now, note the system evolves in a four-dimensional (4D) subspace, in contrast to the first-order Kronecker graph (complete graph) that evolves in a 2D subspace \cite{CG2004} and the second-order Kronecker graph (strongly regular graph) that evolves in a 3D subspace \cite{Wong5}. This is because vertices can evolve identically to each other due to the symmetry of the graph and quantum walk. In this case, there are four types of vertices. The details are given in the Appendix, but we briefly describe them here. The first type of vertex is the marked vertex, which evolves uniquely. The second type are the vertices adjacent to the marked vertex. They evolve identically to each other, and there are $(M-1)^3$ of them. The graph has diameter $2$, so vertices nonadjacent to the marked vertex constitute the third and fourth types of vertices. Specifically, $3(M-1)$ of these vertices share $(M-1)^2(M-2)$ common neighbors with the marked vertex, and $3(M-1)^2$ of them share $(M-1)(M-2)^2$ mutual neighbors with the marked vertex. Altogether, there are $1 + (M-1)^3 + 3(M-1) + 3(M-1)^3 = M^3 = N$ vertices, as expected. Each vertex is color-coded by type in Fig.~\ref{fig:K4K4K4_v1_subspace}, with the marked vertex red, its adjacent vertices blue, and the two types of nonadjacent vertices yellow and magenta. Since the graph is vertex transitive, without loss of generality, vertex $1$ can be considered to be the marked vertex.

Grouping together identically-evolving vertices, the 4D subspace is spanned by
\begin{align*}
	\ket{a} &= \ket{\text{Type 1}} = \ket{w}, \\
	\ket{b} &= \frac{1}{\sqrt{(M-1)^3}} \sum_{x \in \text{Type 2}} \ket{x}, \\
	\ket{c} &= \frac{1}{\sqrt{3(M-1)}} \sum_{x \in \text{Type 3}} \ket{x}, \\
	\ket{d} &= \frac{1}{\sqrt{3(M-1)^2}} \sum_{x \in \text{Type 4}} \ket{x}.
\end{align*}
So $\ket{a}$ is the marked vertex, $\ket{b}$ are the vertices adjacent to the marked vertex, and $\ket{c}$ and $\ket{d}$ are the two types of vertices nonadjacent to the marked vertex. In this $\{ \ket{a}, \ket{b}, \ket{c}, \ket{d} \}$ basis, the initial uniform superposition state \eqref{eq:s} is
\[ \ket{s} = \frac{1}{\sqrt{N}} \begin{pmatrix}
	1 \\
	\sqrt{(M-1)^3} \\
	\sqrt{3(M-1)} \\
	\sqrt{3(M-1)^2} \\
\end{pmatrix}, \]
and the adjacency matrix is
\[ A^{\otimes 3} = \begin{pmatrix}
	0 & \sqrt{M_1^3} & 0 & 0 \\
	\sqrt{M_1^3} & M_2^3 & \sqrt{3}M_1M_2 & \sqrt{3M_1} M_2^2 \\
	0 & \sqrt{3}M_1M_2 & 0 & \sqrt{M_1^3} \\
	0 & \sqrt{3M_1}M_2^2 & \sqrt{M_1^3} & 2M_1M_2 \\
\end{pmatrix}, \]
where $M_i = M-i$. For example, the entry in the third row, second column of $A$ comes from the $(M-1)^2(M-2)$ Type 2 vertices adjacent to a Type 3 vertex, times $\sqrt{3(M-1)}$ and divided by $\sqrt{(M-1)^3}$ to convert between the normalizations of $\ket{c}$ and $\ket{b}$. Using this, the search Hamiltonian \eqref{eq:H} is
\[ H = -\gamma \begin{pmatrix}
	\frac{1}{\gamma} & \sqrt{M_1^3} & 0 & 0 \\
	\sqrt{M_1^3} & M_2^3 & \sqrt{3}M_1M_2 & \sqrt{3M_1} M_2^2 \\
	0 & \sqrt{3}M_1M_2 & 0 & \sqrt{M_1^3} \\
	0 & \sqrt{3M_1}M_2^2 & \sqrt{M_1^3} & 2M_1M_2 \\
\end{pmatrix}. \]

To determine how the search algorithm evolves with this Hamiltonian for large $N$, we utilize degenerate perturbation theory. In this approach \cite{Wong5,Wong8}, we first decompose the Hamiltonian into leading- and higher-order terms:
\[ H = H^{(0)} + H^{(1)} + \dots \]
for large $M$. From this, we next find the eigenvalues and eigenvectors of $H^{(0)}$, some of which may be degenerate. Finally, adding the perturbation $H^{(1)}$, certain linear combinations of the degenerate eigenvectors of $H^{(0)}$ are eigenvectors of $H^{(0)} + H^{(1)}$, and this ``lifts'' the degeneracy. This mixing drives evolution between degenerate eigenvectors of $H^{(0)}$, and the energy or eigenvalue gap dictates the rate of evolution.

We want the perturbation $H^{(1)}$ to mix the marked vertex $\ket{a}$ with the unmarked vertices and drive evolution between them, and this only occurs from the $\bra{a} H \ket{b} = \bra{b} H \ket{a} = \sqrt{(M-1)^3}$ terms, which is $O(M^{3/2})$, apart from a factor of $-\gamma$. So, $H^{(1)}$ should include terms $\Theta(M^{3/2})$, and $H^{(0)}$ should include anything of higher order, i.e., $\omega(M^{3/2})$:
\[ H^{(0)} = -\gamma \begin{pmatrix}
	\frac{1}{\gamma} & 0 & 0 & 0 \\
	0 & M^3 - 6M^2 & \sqrt{3} M^2 & \sqrt{3} M^{5/2} \\
	0 & \sqrt{3} M^2 & 0 & 0 \\
	0 & \sqrt{3} M^{5/2} & 0 & 2M^2 \\
\end{pmatrix}, \]
\[ H^{(1)} = -\gamma \begin{pmatrix}
	0 & M^{3/2} & 0 & 0 \\
	M^{3/2} & 0 & 0 & -\frac{9\sqrt{3}}{2} M^{3/2} \\
	0 & 0 & 0 & M^{3/2} \\
	0 & -\frac{9\sqrt{3}}{2} M^{3/2} & M^{3/2} & 0 \\
\end{pmatrix}, \]
\[ H^{(2)} = O(\gamma M). \]
With this decomposition, we next need to find the eigenvectors and eigenvalues of $H^{(0)}$, but unfortunately, this is prohibitively complicated.

To circumvent this obstacle, we can try changing the basis, as in \cite{Wong5,Wong20}. Besides $\ket{a}$, we choose the uniform superposition of unmarked vertices to be another basis state:
\begin{widetext}
\[ \ket{r} = \frac{1}{\sqrt{N-1}} \sum_{x \ne w} \ket{x} = \frac{1}{\sqrt{M^3-1}} \Big( \sqrt{(M-1)^3} \ket{b} + \sqrt{3(M-1)} \ket{c} + \sqrt{3(M-1)^2} \ket{d} \Big). \]
A state that is obviously orthogonal to this, which we use as a third basis state, is
\[ \ket{r'} = \frac{1}{\sqrt{M}} \left( \sqrt{M-1} \ket{c} - \ket{d} \right). \]
For the fourth basis state, we take the cross product of $\ket{r}$ and $\ket{r'}$. Abusing notation,
\begin{align*}
	\ket{r''} 
		&= \ket{r} \times \ket{r'} \\
		&= \frac{1}{\sqrt{M^3-1}} \frac{1}{\sqrt{M}} \begin{vmatrix}
			\ket{b} & \ket{c} & \ket{d} \\
			\sqrt{(M-1)^3} & \sqrt{3(M-1)} & \sqrt{3(M-1)^2} \\
			0 & \sqrt{M-1} & -1 \\
		\end{vmatrix} \\
		&= \frac{1}{\sqrt{M^3-1}} \frac{1}{\sqrt{M}} \left( -M\sqrt{3(M-1)} \ket{b} + \sqrt{(M-1)^3} \ket{c} + (M-1)^2 \ket{d} \right).
\end{align*}
Changing the Hamiltonian from the $\{ \ket{a}, \ket{b}, \ket{c}, \ket{d} \}$ basis to the $\{ \ket{a}, \ket{r}, \ket{r'}, \ket{r''} \}$ basis, we conjugate it by
\[ T = \begin{pmatrix}
	\ket{a} & \ket{r} & \ket{r'} & \ket{r''} \\
\end{pmatrix}. \]
That is, we calculate $T^{-1} H T$ to get the Hamiltonian in the new basis. Note that in this case, $T^{-1} = T^\intercal$. Doing this and keeping terms at least linear in $M$, the Hamiltonian in the $\{ \ket{a}, \ket{r}, \ket{r'}, \ket{r''} \}$ basis is
\[ H' \approx -\gamma \begin{pmatrix}
	\frac{1}{\gamma} & M^{3/2} & 0 & -\sqrt{3} M \\
	M^{3/2} & M^3 -3M^2 + 3M & 0 & 0 \\
	0 & 0 & 0 & -M^{3/2} \\
	-\sqrt{3} M & 0 & -M^{3/2} & -M^2 + 3M \\
\end{pmatrix}. \]
\end{widetext}

Utilizing degenerate perturbation theory in this new basis, we decompose the Hamiltonian $H'$ into leading- and higher-order matrices:
\begin{gather*}
H'^{(0)} = -\gamma \begin{pmatrix}
	\frac{1}{\gamma} & 0 & 0 & 0 \\
	0 & M^3 - 3M^2 & 0 & 0 \\
	0 & 0 & 0 & 0 \\
	0 & 0 & 0 & -M^2 \\
\end{pmatrix}, \\
H'^{(1)} = -\gamma \begin{pmatrix}
	0 & M^{3/2} & 0 & 0 \\
	M^{3/2} & 0 & 0 & 0 \\
	0 & 0 & 0 & -M^{3/2} \\
	0 & 0 & -M^{3/2} & 0 \\
\end{pmatrix}.
\end{gather*}
The eigenvectors and corresponding eigenvalues of $H'^{(0)}$ are now easy to identify, unlike in the old basis. They are
\begin{align*}
	&\ket{a}, \quad -1 \\
	&\ket{r}, \quad -\gamma (M^3 - 3M^2) \\
	&\ket{r'}, \quad 0 \\
	&\ket{r''}, \quad \gamma M^2. 
\end{align*}
The initial state $\ket{s}$ is approximately $\ket{r}$ for large $M$, and since we want this to evolve to $\ket{a}$, we make $\ket{a}$ and $\ket{r}$ degenerate by choosing:
\begin{equation}
	\label{eq:gamma}
	\gamma = \frac{1}{M^2(M-3)}.
\end{equation}
This is the ``critical value'' of $\gamma$.

\begin{figure}
\begin{center}
	\includegraphics{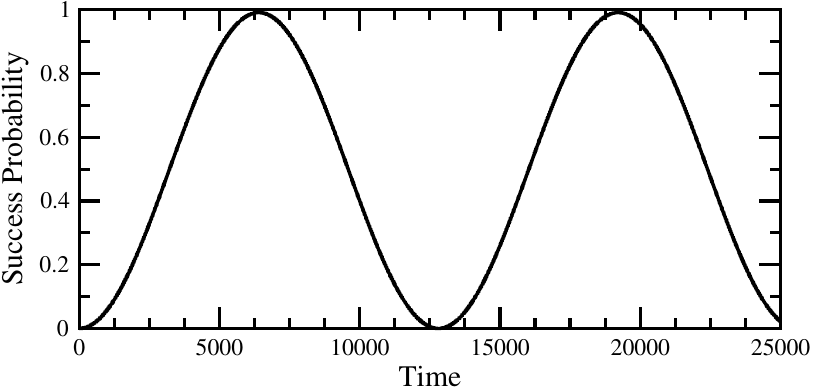}
	\caption{\label{fig:prob_time_j3} Success probability as a function of time for search by continuous-time quantum walk on $K_M \otimes K_M \otimes K_M$ with $M = 256$.}
\end{center}
\end{figure}

Next, we include the perturbation $H'^{(1)}$, which causes linear combinations $\alpha_a \ket{a} + \alpha_r \ket{r}$ to be eigenvectors of $H'^{(0)} + H'^{(1)}$. To find the coefficients, we solve the eigenvalue problem
\[ \begin{pmatrix}
	H_{aa} & H_{ar} \\
	H_{ra} & H_{rr} \\
\end{pmatrix} \begin{pmatrix}
	\alpha_a \\
	\alpha_r \\
\end{pmatrix} = E \begin{pmatrix}
	\alpha_a \\
	\alpha_r \\
\end{pmatrix}, \]
where $H_{ar} = \bra{a} (H'^{(0)} + H'^{(1)}) \ket{r}$, etc., and $E$ is the eigenvalue. Evaluating the matrix elements with $\gamma$ at its critical value \eqref{eq:gamma},
\[ \begin{pmatrix}
	-1 & \frac{-1}{\sqrt{M}(M-3)} \\
	\frac{-1}{\sqrt{M}(M-3)} & -1 \\
\end{pmatrix} \begin{pmatrix}
	\alpha_a \\
	\alpha_r \\
\end{pmatrix} = E \begin{pmatrix}
	\alpha_a \\
	\alpha_r \\
\end{pmatrix}. \]
Solving this, we get the following eigenvectors and eigenvalues of $H'^{(0)} + H'^{(1)}$:
\[ \begin{pmatrix}
	\alpha_a \\
	\alpha_r \\
\end{pmatrix} = \frac{1}{\sqrt{2}} \begin{pmatrix}
	1 \\
	1 \\
\end{pmatrix}, \quad E_0 = -1 - \frac{1}{\sqrt{M}(M-3)}
\]
\[ \begin{pmatrix}
	\alpha_a \\
	\alpha_r \\
\end{pmatrix} = \frac{1}{\sqrt{2}} \begin{pmatrix}
	-1 \\
	1 \\
\end{pmatrix}, \quad E_1 = -1 + \frac{1}{\sqrt{M}(M-3)}.
\]
Since the eigenvectors are proportional to $\pm \ket{a} + \ket{r}$, the system evolves from $\ket{s} \approx \ket{r}$ to $\ket{a}$, up to a phase, in time
\[ \frac{\pi}{\Delta E} = \frac{\pi}{2} \sqrt{M}(M-3) \approx \frac{\pi}{2} M^{3/2} \approx \frac{\pi}{2} \sqrt{N}, \]
where $\Delta E = E_1 - E_0$ is the energy gap. This is the same runtime as the first-order Kronecker graph (i.e., complete graph) and second-order Kronecker graph. A simulation is shown in Fig.~\ref{fig:prob_time_j3}, and the success probability reaches $1$ at time $\pi\sqrt{256^3}/2 \approx 6433.98$, as expected.

These results are asymptotic, meaning $M$ must be sufficiently large in order for the success probability to reach $1$ at time $\pi\sqrt{N}/2$. Numerically, choosing $\gamma$ to be $1/(M-1)^3$, rather than $1/M^2(M-3)$ that we derived earlier, allows smaller values of $M$ to exhibit this asymptotic runtime. Taylor expanding the two values of $\gamma$ for large $M$,
\begin{gather*}
	\frac{1}{(M-1)^3} = \frac{1}{M^3} + \frac{3}{M^4} + \frac{6}{M^5} + O\left( \frac{1}{M^6} \right), \\
	\frac{1}{M^2(M-3)} = \frac{1}{M^3} + \frac{3}{M^4} + \frac{9}{M^5} + O\left( \frac{1}{M^6} \right).
\end{gather*}
Thus, the two values of $\gamma$ are equal, up to terms of order $1/M^5$.


\section{Higher Orders}

\begin{figure}
\begin{center}
	\includegraphics{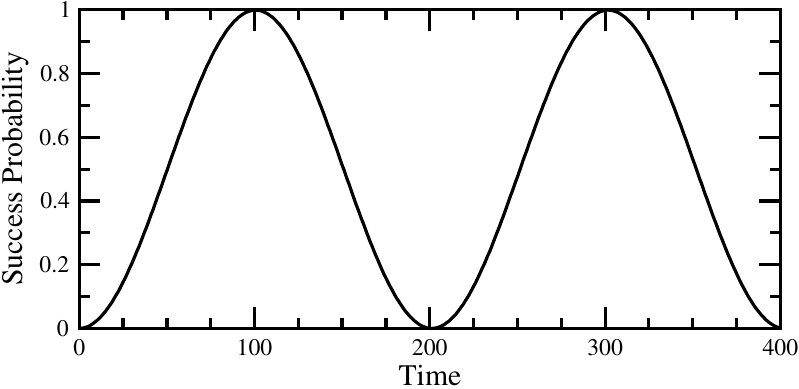}
	\caption{\label{fig:prob_time_j6} Success probability as a function of time for search by continuous-time quantum walk on the sixth-order Kronecker graph $K_M \otimes K_M \otimes K_M \otimes K_M \otimes K_M \otimes K_M$ with $M = 4$ and $\gamma = 0.001372$.}
\end{center}
\end{figure}

In theory, we can apply the same perturbative approach from the third-order case to Kronecker graphs with order $j \ge 4$. Analyzing an infinite number of these cases in this iterative manner, however, is prohibitive due to the countably infinite possible values for $j$. Further complicating the matter is that the dimension of the subspace increases with $j$, so perturbation theory would be used to find the eigenvalues and eigenvectors of progressively larger matrices, making generalization of our results difficult. For example, when $j = 1$, the complete graph evolves in a 2D subspace, when $j = 2$, the strongly regular graph evolves in a 3D subspace, and when $j = 3$, the graph evolves in a 4D subspace.

Numerical simulations do suggest, however, that search higher-order Kronecker graphs with the complete initiator also takes $\pi\sqrt{N}/2$ time, asymptotically. For example, the success probability for search on the sixth-order Kronecker graph is shown in Fig.~\ref{fig:prob_time_j6}, and it reaches $1$ at time $\pi\sqrt{4^6}/2 \approx 100.53$. Similarly, numerical simulations show that increasing $j$ and keeping $M$ fixed, or increasing both $j$ and $M$, yields a successful search in $\pi\sqrt{M^j}/2$ time.

Despite the increase in the dimension of the subspace, we can prove that the diameter of the Kronecker graph, with complete initiator, is $2$ for all $j \ge 2$ and $M \ge 3$. To do this, note a general vertex $\ket{v}$ can be written as
\[ \ket{v} = \ket{p_1,p_2, \dots, p_j}, \]
where each $p_i$ takes values $1, \dots, M$ and encodes the position at each level of hierarchy. For example, in Fig.~\ref{fig:prob_time_j6}, if $p_i = 1$ encodes the top-left position, then vertex $1$ in would be $\ket{1,1,1}$ since it is the top-left vertex in its subset $P_{1,1}$, its subset $P_{1,1}$ is the top-left subset in its set $P_1$, and $P_1$ is the top-left set. A vertex nonadjacent to $\ket{v}$ has the general form
\[ \ket{u} = \ket{p_1',p_2', \dots, p_j'}, \]
where at least one $p_i' = p_i$ since, if all of the positions were to differ, the vertices would be adjacent. Now consider a third vertex
\[ \ket{w} = \ket{p_1'',p_2'', \dots, p_j''}, \]
where $p_i'' \ne p_i$ and $p_i'' \ne p_i'$. This is possible since $M \ge 3$. Then, $\ket{w}$ is adjacent to both $\ket{v}$ and $\ket{u}$, so the distance between $\ket{v}$ and $\ket{u}$ is $2$, and the diameter of the graph is $2$.


\section{Conclusion}

Kronecker graphs are a useful method to generate complex networks with real-world properties. Here, we investigated how continuous-time quantum walks search Kronecker graphs by focusing on the case that the initiator is the complete graph. Then, the first-order Kronecker graph is exactly the quantum walk formulation of Grover's algorithm, where the search takes $\pi\sqrt{N}/2$ time. We proved that second-order Kronecker graph is a strongly regular graph, and that it is also searched in $\pi\sqrt{N}/2$ time. Furthermore, using degenerate perturbation theory, we proved that the third-order Kronecker graph is searched in the same optimal runtime as the first- and second-order graphs. Numerical simulations indicate that higher-order Kronecker graphs behave the same way, and an analytical proof is open for further research. Our work on quantum search on Kronecker graphs, where the initiator is not the complete graph, will be reported elsewhere.


\begin{acknowledgments}
	Thanks to Chris Godsil for useful discussions. This project is a part of K.W.'s Master's thesis. J.L.~acknowledges financial support by the Engineering and Physical Sciences Research Council [grant number EP/L015242/1]. S.S.~is supported by the Royal Society, EPSRC, the National Natural Science Foundation of China, and the grant ARO-MURI W911NF-17-1-0304 (US DOD, UK MOD, and UK EPSRC under the Multidisciplinary University Research Initiative).
\end{acknowledgments}


\bibliography{refs}

\begin{thebibliography}{27}%
\makeatletter
\providecommand \@ifxundefined [1]{%
 \@ifx{#1\undefined}
}%
\providecommand \@ifnum [1]{%
 \ifnum #1\expandafter \@firstoftwo
 \else \expandafter \@secondoftwo
 \fi
}%
\providecommand \@ifx [1]{%
 \ifx #1\expandafter \@firstoftwo
 \else \expandafter \@secondoftwo
 \fi
}%
\providecommand \natexlab [1]{#1}%
\providecommand \enquote  [1]{``#1''}%
\providecommand \bibnamefont  [1]{#1}%
\providecommand \bibfnamefont [1]{#1}%
\providecommand \citenamefont [1]{#1}%
\providecommand \href@noop [0]{\@secondoftwo}%
\providecommand \href [0]{\begingroup \@sanitize@url \@href}%
\providecommand \@href[1]{\@@startlink{#1}\@@href}%
\providecommand \@@href[1]{\endgroup#1\@@endlink}%
\providecommand \@sanitize@url [0]{\catcode `\\12\catcode `\$12\catcode
  `\&12\catcode `\#12\catcode `\^12\catcode `\_12\catcode `\%12\relax}%
\providecommand \@@startlink[1]{}%
\providecommand \@@endlink[0]{}%
\providecommand \url  [0]{\begingroup\@sanitize@url \@url }%
\providecommand \@url [1]{\endgroup\@href {#1}{\urlprefix }}%
\providecommand \urlprefix  [0]{URL }%
\providecommand \Eprint [0]{\href }%
\providecommand \doibase [0]{http://dx.doi.org/}%
\providecommand \selectlanguage [0]{\@gobble}%
\providecommand \bibinfo  [0]{\@secondoftwo}%
\providecommand \bibfield  [0]{\@secondoftwo}%
\providecommand \translation [1]{[#1]}%
\providecommand \BibitemOpen [0]{}%
\providecommand \bibitemStop [0]{}%
\providecommand \bibitemNoStop [0]{.\EOS\space}%
\providecommand \EOS [0]{\spacefactor3000\relax}%
\providecommand \BibitemShut  [1]{\csname bibitem#1\endcsname}%
\let\auto@bib@innerbib\@empty
\bibitem [{\citenamefont {Grover}(1996)}]{Grover1996}%
  \BibitemOpen
  \bibfield  {author} {\bibinfo {author} {\bibfnamefont {L.~K.}\ \bibnamefont
  {Grover}},\ }\bibfield  {title} {\enquote {\bibinfo {title} {A fast quantum
  mechanical algorithm for database search},}\ }in\ \href@noop {} {\emph
  {\bibinfo {booktitle} {Proceedings of the 28th Annual ACM Symposium on Theory
  of Computing}}},\ \bibinfo {series and number} {STOC '96}\ (\bibinfo
  {publisher} {ACM},\ \bibinfo {address} {New York, NY, USA},\ \bibinfo {year}
  {1996})\ pp.\ \bibinfo {pages} {212--219}\BibitemShut {NoStop}%
\bibitem [{\citenamefont {Nielsen}\ and\ \citenamefont
  {Chuang}(2000)}]{NielsenChuang2000}%
  \BibitemOpen
  \bibfield  {author} {\bibinfo {author} {\bibfnamefont {M.~A.}\ \bibnamefont
  {Nielsen}}\ and\ \bibinfo {author} {\bibfnamefont {I.~L.}\ \bibnamefont
  {Chuang}},\ }\href@noop {} {\emph {\bibinfo {title} {Quantum Computation and
  Quantum Information}}}\ (\bibinfo  {publisher} {Cambridge University Press},\
  \bibinfo {year} {2000})\BibitemShut {NoStop}%
\bibitem [{\citenamefont {Benioff}(2002)}]{Benioff2002}%
  \BibitemOpen
  \bibfield  {author} {\bibinfo {author} {\bibfnamefont {P.}~\bibnamefont
  {Benioff}},\ }\enquote {\bibinfo {title} {Space searches with a quantum
  robot},}\ in\ \href {\doibase 10.1090/conm/305} {\emph {\bibinfo {booktitle}
  {Quantum Computation and Information}}},\ \bibinfo {series} {Contemp. Math.},
  Vol.\ \bibinfo {volume} {305}\ (\bibinfo  {publisher} {Amer. Math. Soc.},\
  \bibinfo {address} {Providence, RI},\ \bibinfo {year} {2002})\ pp.\ \bibinfo
  {pages} {1--12}\BibitemShut {NoStop}%
\bibitem [{\citenamefont {Meyer}(1996{\natexlab{a}})}]{Meyer1996a}%
  \BibitemOpen
  \bibfield  {author} {\bibinfo {author} {\bibfnamefont {D.~A.}\ \bibnamefont
  {Meyer}},\ }\bibfield  {title} {\enquote {\bibinfo {title} {From quantum
  cellular automata to quantum lattice gases},}\ }\href {\doibase
  10.1007/BF02199356} {\bibfield  {journal} {\bibinfo  {journal} {J. Stat.
  Phys.}\ }\textbf {\bibinfo {volume} {85}},\ \bibinfo {pages} {551--574}
  (\bibinfo {year} {1996}{\natexlab{a}})}\BibitemShut {NoStop}%
\bibitem [{\citenamefont {Meyer}(1996{\natexlab{b}})}]{Meyer1996b}%
  \BibitemOpen
  \bibfield  {author} {\bibinfo {author} {\bibfnamefont {D.~A.}\ \bibnamefont
  {Meyer}},\ }\bibfield  {title} {\enquote {\bibinfo {title} {On the absence of
  homogeneous scalar unitary cellular automata},}\ }\href {\doibase
  http://dx.doi.org/10.1016/S0375-9601(96)00745-1} {\bibfield  {journal}
  {\bibinfo  {journal} {Phys. Lett. A}\ }\textbf {\bibinfo {volume} {223}},\
  \bibinfo {pages} {337--340} (\bibinfo {year}
  {1996}{\natexlab{b}})}\BibitemShut {NoStop}%
\bibitem [{\citenamefont {Aaronson}\ and\ \citenamefont
  {Ambainis}(2005)}]{AA2005}%
  \BibitemOpen
  \bibfield  {author} {\bibinfo {author} {\bibfnamefont {S.}~\bibnamefont
  {Aaronson}}\ and\ \bibinfo {author} {\bibfnamefont {A.}~\bibnamefont
  {Ambainis}},\ }\bibfield  {title} {\enquote {\bibinfo {title} {Quantum search
  of spatial regions},}\ }\href {\doibase 10.4086/toc.2005.v001a004} {\bibfield
   {journal} {\bibinfo  {journal} {Theor. Comput.}\ }\textbf {\bibinfo {volume}
  {1}},\ \bibinfo {pages} {47--79} (\bibinfo {year} {2005})}\BibitemShut
  {NoStop}%
\bibitem [{\citenamefont {Childs}\ and\ \citenamefont
  {Goldstone}(2004)}]{CG2004}%
  \BibitemOpen
  \bibfield  {author} {\bibinfo {author} {\bibfnamefont {A.~M.}\ \bibnamefont
  {Childs}}\ and\ \bibinfo {author} {\bibfnamefont {J.}~\bibnamefont
  {Goldstone}},\ }\bibfield  {title} {\enquote {\bibinfo {title} {Spatial
  search by quantum walk},}\ }\href {\doibase 10.1103/PhysRevA.70.022314}
  {\bibfield  {journal} {\bibinfo  {journal} {Phys. Rev. A}\ }\textbf {\bibinfo
  {volume} {70}},\ \bibinfo {pages} {022314} (\bibinfo {year}
  {2004})}\BibitemShut {NoStop}%
\bibitem [{\citenamefont {Ambainis}\ \emph {et~al.}(2005)\citenamefont
  {Ambainis}, \citenamefont {Kempe},\ and\ \citenamefont {Rivosh}}]{AKR2005}%
  \BibitemOpen
  \bibfield  {author} {\bibinfo {author} {\bibfnamefont {A.}~\bibnamefont
  {Ambainis}}, \bibinfo {author} {\bibfnamefont {J.}~\bibnamefont {Kempe}}, \
  and\ \bibinfo {author} {\bibfnamefont {A.}~\bibnamefont {Rivosh}},\
  }\bibfield  {title} {\enquote {\bibinfo {title} {Coins make quantum walks
  faster},}\ }in\ \href@noop {} {\emph {\bibinfo {booktitle} {Proceedings of
  the 16th Annual ACM-SIAM Symposium on Discrete Algorithms}}},\ \bibinfo
  {series and number} {SODA '05}\ (\bibinfo  {publisher} {SIAM},\ \bibinfo
  {address} {Philadelphia, PA, USA},\ \bibinfo {year} {2005})\ pp.\ \bibinfo
  {pages} {1099--1108}\BibitemShut {NoStop}%
\bibitem [{\citenamefont {Janmark}\ \emph {et~al.}(2014)\citenamefont
  {Janmark}, \citenamefont {Meyer},\ and\ \citenamefont {Wong}}]{Wong5}%
  \BibitemOpen
  \bibfield  {author} {\bibinfo {author} {\bibfnamefont {J.}~\bibnamefont
  {Janmark}}, \bibinfo {author} {\bibfnamefont {D.~A.}\ \bibnamefont {Meyer}},
  \ and\ \bibinfo {author} {\bibfnamefont {T.~G.}\ \bibnamefont {Wong}},\
  }\bibfield  {title} {\enquote {\bibinfo {title} {Global symmetry is
  unnecessary for fast quantum search},}\ }\href {\doibase
  10.1103/PhysRevLett.112.210502} {\bibfield  {journal} {\bibinfo  {journal}
  {Phys. Rev. Lett.}\ }\textbf {\bibinfo {volume} {112}},\ \bibinfo {pages}
  {210502} (\bibinfo {year} {2014})}\BibitemShut {NoStop}%
\bibitem [{\citenamefont {Meyer}\ and\ \citenamefont {Wong}(2015)}]{Wong7}%
  \BibitemOpen
  \bibfield  {author} {\bibinfo {author} {\bibfnamefont {D.~A.}\ \bibnamefont
  {Meyer}}\ and\ \bibinfo {author} {\bibfnamefont {T.~G.}\ \bibnamefont
  {Wong}},\ }\bibfield  {title} {\enquote {\bibinfo {title} {Connectivity is a
  poor indicator of fast quantum search},}\ }\href {\doibase
  10.1103/PhysRevLett.114.110503} {\bibfield  {journal} {\bibinfo  {journal}
  {Phys. Rev. Lett.}\ }\textbf {\bibinfo {volume} {114}},\ \bibinfo {pages}
  {110503} (\bibinfo {year} {2015})}\BibitemShut {NoStop}%
\bibitem [{\citenamefont {Chakraborty}\ \emph {et~al.}(2016)\citenamefont
  {Chakraborty}, \citenamefont {Novo}, \citenamefont {Ambainis},\ and\
  \citenamefont {Omar}}]{Chakraborty2016}%
  \BibitemOpen
  \bibfield  {author} {\bibinfo {author} {\bibfnamefont {S.}~\bibnamefont
  {Chakraborty}}, \bibinfo {author} {\bibfnamefont {L.}~\bibnamefont {Novo}},
  \bibinfo {author} {\bibfnamefont {A.}~\bibnamefont {Ambainis}}, \ and\
  \bibinfo {author} {\bibfnamefont {Y.}~\bibnamefont {Omar}},\ }\bibfield
  {title} {\enquote {\bibinfo {title} {Spatial search by quantum walk is
  optimal for almost all graphs},}\ }\href {\doibase
  10.1103/PhysRevLett.116.100501} {\bibfield  {journal} {\bibinfo  {journal}
  {Phys. Rev. Lett.}\ }\textbf {\bibinfo {volume} {116}},\ \bibinfo {pages}
  {100501} (\bibinfo {year} {2016})}\BibitemShut {NoStop}%
\bibitem [{\citenamefont {Kempe}(2003)}]{Kempe2003}%
  \BibitemOpen
  \bibfield  {author} {\bibinfo {author} {\bibfnamefont {J.}~\bibnamefont
  {Kempe}},\ }\bibfield  {title} {\enquote {\bibinfo {title} {Quantum random
  walks: An introductory overview},}\ }\href@noop {} {\bibfield  {journal}
  {\bibinfo  {journal} {Contemp. Phys.}\ }\textbf {\bibinfo {volume} {44}},\
  \bibinfo {pages} {307--327} (\bibinfo {year} {2003})}\BibitemShut {NoStop}%
\bibitem [{\citenamefont {Milgram}(1967)}]{Milgram1967}%
  \BibitemOpen
  \bibfield  {author} {\bibinfo {author} {\bibfnamefont {S.}~\bibnamefont
  {Milgram}},\ }\bibfield  {title} {\enquote {\bibinfo {title} {The small-world
  problem},}\ }\href@noop {} {\bibfield  {journal} {\bibinfo  {journal}
  {Psychology Today}\ }\textbf {\bibinfo {volume} {1}},\ \bibinfo {pages}
  {61--69} (\bibinfo {year} {1967})}\BibitemShut {NoStop}%
\bibitem [{\citenamefont {Barab{\'a}si}\ and\ \citenamefont
  {Albert}(1999)}]{Barabasi1999}%
  \BibitemOpen
  \bibfield  {author} {\bibinfo {author} {\bibfnamefont {A.-L.}\ \bibnamefont
  {Barab{\'a}si}}\ and\ \bibinfo {author} {\bibfnamefont {R.}~\bibnamefont
  {Albert}},\ }\bibfield  {title} {\enquote {\bibinfo {title} {Emergence of
  scaling in random networks},}\ }\href {\doibase 10.1126/science.286.5439.509}
  {\bibfield  {journal} {\bibinfo  {journal} {Science}\ }\textbf {\bibinfo
  {volume} {286}},\ \bibinfo {pages} {509--512} (\bibinfo {year}
  {1999})}\BibitemShut {NoStop}%
\bibitem [{\citenamefont {Leskovec}\ and\ \citenamefont
  {Krevl}(2014)}]{snapnets}%
  \BibitemOpen
  \bibfield  {author} {\bibinfo {author} {\bibfnamefont {J.}~\bibnamefont
  {Leskovec}}\ and\ \bibinfo {author} {\bibfnamefont {A.}~\bibnamefont
  {Krevl}},\ }\href@noop {} {\enquote {\bibinfo {title} {{SNAP Datasets}:
  {Stanford} large network dataset collection},}\ }\bibinfo {howpublished}
  {\url{http://snap.stanford.edu/data}} (\bibinfo {year} {2014})\BibitemShut
  {NoStop}%
\bibitem [{\citenamefont {Stumpf}\ \emph {et~al.}(2005)\citenamefont {Stumpf},
  \citenamefont {Wiuf},\ and\ \citenamefont {May}}]{Stumpf2005}%
  \BibitemOpen
  \bibfield  {author} {\bibinfo {author} {\bibfnamefont {M.~P.~H.}\
  \bibnamefont {Stumpf}}, \bibinfo {author} {\bibfnamefont {C.}~\bibnamefont
  {Wiuf}}, \ and\ \bibinfo {author} {\bibfnamefont {R.~M.}\ \bibnamefont
  {May}},\ }\bibfield  {title} {\enquote {\bibinfo {title} {Subnets of
  scale-free networks are not scale-free: Sampling properties of networks},}\
  }\href {\doibase 10.1073/pnas.0501179102} {\bibfield  {journal} {\bibinfo
  {journal} {Proc. Natl. Acad. Sci. U.S.A.}\ }\textbf {\bibinfo {volume}
  {102}},\ \bibinfo {pages} {4221--4224} (\bibinfo {year} {2005})}\BibitemShut
  {NoStop}%
\bibitem [{\citenamefont {Leskovec}\ \emph
  {et~al.}(2005{\natexlab{a}})\citenamefont {Leskovec}, \citenamefont
  {Kleinberg},\ and\ \citenamefont {Faloutsos}}]{Leskovec2005a}%
  \BibitemOpen
  \bibfield  {author} {\bibinfo {author} {\bibfnamefont {J.}~\bibnamefont
  {Leskovec}}, \bibinfo {author} {\bibfnamefont {J.}~\bibnamefont {Kleinberg}},
  \ and\ \bibinfo {author} {\bibfnamefont {C.}~\bibnamefont {Faloutsos}},\
  }\bibfield  {title} {\enquote {\bibinfo {title} {Graphs over time:
  Densification laws, shrinking diameters and possible explanations},}\ }in\
  \href {\doibase 10.1145/1081870.1081893} {\emph {\bibinfo {booktitle} {Proc.
  11th ACM SIGKDD International Conference on Knowledge Discovery in Data
  Mining}}},\ \bibinfo {series and number} {KDD '05}\ (\bibinfo  {publisher}
  {ACM},\ \bibinfo {address} {New York, NY, USA},\ \bibinfo {year} {2005})\
  pp.\ \bibinfo {pages} {177--187}\BibitemShut {NoStop}%
\bibitem [{\citenamefont {Leskovec}\ \emph
  {et~al.}(2005{\natexlab{b}})\citenamefont {Leskovec}, \citenamefont
  {Chakrabarti}, \citenamefont {Kleinberg},\ and\ \citenamefont
  {Faloutsos}}]{Leskovec2005b}%
  \BibitemOpen
  \bibfield  {author} {\bibinfo {author} {\bibfnamefont {J.}~\bibnamefont
  {Leskovec}}, \bibinfo {author} {\bibfnamefont {D.}~\bibnamefont
  {Chakrabarti}}, \bibinfo {author} {\bibfnamefont {J.}~\bibnamefont
  {Kleinberg}}, \ and\ \bibinfo {author} {\bibfnamefont {C.}~\bibnamefont
  {Faloutsos}},\ }\bibfield  {title} {\enquote {\bibinfo {title} {Realistic,
  mathematically tractable graph generation and evolution, using kronecker
  multiplication},}\ }in\ \href {\doibase 10.1007/11564126_17} {\emph {\bibinfo
  {booktitle} {Proc. 9th European Conference on Principles and Practice of
  Knowledge Discovery in Databases}}},\ \bibinfo {series and number} {PKDD
  2005}\ (\bibinfo  {publisher} {Springer},\ \bibinfo {address} {Berlin,
  Heidelberg},\ \bibinfo {year} {2005})\ pp.\ \bibinfo {pages}
  {133--145}\BibitemShut {NoStop}%
\bibitem [{\citenamefont {Leskovec}\ and\ \citenamefont
  {Faloutsos}(2007)}]{Leskovec2007}%
  \BibitemOpen
  \bibfield  {author} {\bibinfo {author} {\bibfnamefont {J.}~\bibnamefont
  {Leskovec}}\ and\ \bibinfo {author} {\bibfnamefont {C.}~\bibnamefont
  {Faloutsos}},\ }\bibfield  {title} {\enquote {\bibinfo {title} {Scalable
  modeling of real graphs using kronecker multiplication},}\ }in\ \href
  {\doibase 10.1145/1273496.1273559} {\emph {\bibinfo {booktitle} {Proc. 24th
  International Conference on Machine Learning}}},\ \bibinfo {series and
  number} {ICML '07}\ (\bibinfo  {publisher} {ACM},\ \bibinfo {address} {New
  York, NY, USA},\ \bibinfo {year} {2007})\ pp.\ \bibinfo {pages}
  {497--504}\BibitemShut {NoStop}%
\bibitem [{\citenamefont {Leskovec}\ \emph {et~al.}(2010)\citenamefont
  {Leskovec}, \citenamefont {Chakrabarti}, \citenamefont {Kleinberg},
  \citenamefont {Faloutsos},\ and\ \citenamefont {Ghahramani}}]{Leskovec2010}%
  \BibitemOpen
  \bibfield  {author} {\bibinfo {author} {\bibfnamefont {J.}~\bibnamefont
  {Leskovec}}, \bibinfo {author} {\bibfnamefont {D.}~\bibnamefont
  {Chakrabarti}}, \bibinfo {author} {\bibfnamefont {J.}~\bibnamefont
  {Kleinberg}}, \bibinfo {author} {\bibfnamefont {C.}~\bibnamefont
  {Faloutsos}}, \ and\ \bibinfo {author} {\bibfnamefont {Z.}~\bibnamefont
  {Ghahramani}},\ }\bibfield  {title} {\enquote {\bibinfo {title} {Kronecker
  graphs: An approach to modeling networks},}\ }\href@noop {} {\bibfield
  {journal} {\bibinfo  {journal} {J. Mach. Learn. Res.}\ }\textbf {\bibinfo
  {volume} {11}},\ \bibinfo {pages} {985--1042} (\bibinfo {year}
  {2010})}\BibitemShut {NoStop}%
\bibitem [{\citenamefont {Brassard}\ \emph {et~al.}(2002)\citenamefont
  {Brassard}, \citenamefont {H\o{}yer}, \citenamefont {Mosca},\ and\
  \citenamefont {Tapp}}]{BHMT2000}%
  \BibitemOpen
  \bibfield  {author} {\bibinfo {author} {\bibfnamefont {G.}~\bibnamefont
  {Brassard}}, \bibinfo {author} {\bibfnamefont {P.}~\bibnamefont {H\o{}yer}},
  \bibinfo {author} {\bibfnamefont {M.}~\bibnamefont {Mosca}}, \ and\ \bibinfo
  {author} {\bibfnamefont {A.}~\bibnamefont {Tapp}},\ }\enquote {\bibinfo
  {title} {Quantum amplitude amplification and estimation},}\ in\ \href
  {\doibase 10.1090/conm/305} {\emph {\bibinfo {booktitle} {Quantum computation
  and information}}},\ \bibinfo {series} {Contemp. Math.}, Vol.\ \bibinfo
  {volume} {305}\ (\bibinfo  {publisher} {Amer. Math. Soc.},\ \bibinfo
  {address} {Providence, RI},\ \bibinfo {year} {2002})\ pp.\ \bibinfo {pages}
  {53--–74}\BibitemShut {NoStop}%
\bibitem [{\citenamefont {Farhi}\ and\ \citenamefont {Gutmann}(1998)}]{FG1998}%
  \BibitemOpen
  \bibfield  {author} {\bibinfo {author} {\bibfnamefont {E.}~\bibnamefont
  {Farhi}}\ and\ \bibinfo {author} {\bibfnamefont {S.}~\bibnamefont
  {Gutmann}},\ }\bibfield  {title} {\enquote {\bibinfo {title} {Analog analogue
  of a digital quantum computation},}\ }\href {\doibase
  10.1103/PhysRevA.57.2403} {\bibfield  {journal} {\bibinfo  {journal} {Phys.
  Rev. A}\ }\textbf {\bibinfo {volume} {57}},\ \bibinfo {pages} {2403--2406}
  (\bibinfo {year} {1998})}\BibitemShut {NoStop}%
\bibitem [{\citenamefont {Wong}(2015{\natexlab{a}})}]{Wong10}%
  \BibitemOpen
  \bibfield  {author} {\bibinfo {author} {\bibfnamefont {T.~G.}\ \bibnamefont
  {Wong}},\ }\bibfield  {title} {\enquote {\bibinfo {title} {Grover search with
  lackadaisical quantum walks},}\ }\href {\doibase
  10.1088/1751-8113/48/43/435304} {\bibfield  {journal} {\bibinfo  {journal}
  {J. Phys. A: Math. Theor.}\ }\textbf {\bibinfo {volume} {48}},\ \bibinfo
  {pages} {435304} (\bibinfo {year} {2015}{\natexlab{a}})}\BibitemShut
  {NoStop}%
\bibitem [{\citenamefont {Godsil}(2017)}]{Godsil2017}%
  \BibitemOpen
  \bibfield  {author} {\bibinfo {author} {\bibfnamefont {C.}~\bibnamefont
  {Godsil}},\ }\href@noop {} {}\bibinfo {howpublished} {personal communication}
  (\bibinfo {year} {2017})\BibitemShut {NoStop}%
\bibitem [{\citenamefont {Cameron}\ and\ \citenamefont {van
  Lint}(1991)}]{Cameron1991}%
  \BibitemOpen
  \bibfield  {author} {\bibinfo {author} {\bibfnamefont {P.J.}\ \bibnamefont
  {Cameron}}\ and\ \bibinfo {author} {\bibfnamefont {J.H.}\ \bibnamefont {van
  Lint}},\ }\href {http://books.google.com/books?id=j1CZeuHI7q0C} {\emph
  {\bibinfo {title} {Designs, Graphs, Codes and Their Links}}},\ London
  Mathematical Society Student Texts\ (\bibinfo  {publisher} {Cambridge
  University Press},\ \bibinfo {year} {1991})\BibitemShut {NoStop}%
\bibitem [{\citenamefont {Wong}(2015{\natexlab{b}})}]{Wong8}%
  \BibitemOpen
  \bibfield  {author} {\bibinfo {author} {\bibfnamefont {T.~G.}\ \bibnamefont
  {Wong}},\ }\bibfield  {title} {\enquote {\bibinfo {title} {Diagrammatic
  approach to quantum search},}\ }\href {\doibase 10.1007/s11128-015-0959-3}
  {\bibfield  {journal} {\bibinfo  {journal} {Quantum Inf. Process.}\ }\textbf
  {\bibinfo {volume} {14}},\ \bibinfo {pages} {1767--1775} (\bibinfo {year}
  {2015}{\natexlab{b}})}\BibitemShut {NoStop}%
\bibitem [{\citenamefont {Wong}(2016)}]{Wong20}%
  \BibitemOpen
  \bibfield  {author} {\bibinfo {author} {\bibfnamefont {T.~G.}\ \bibnamefont
  {Wong}},\ }\bibfield  {title} {\enquote {\bibinfo {title} {Quantum walk
  search on {J}ohnson graphs},}\ }\href {\doibase
  10.1088/1751-8113/49/19/195303} {\bibfield  {journal} {\bibinfo  {journal}
  {J. Phys. A: Math. Theor.}\ }\textbf {\bibinfo {volume} {49}},\ \bibinfo
  {pages} {195303} (\bibinfo {year} {2016})}\BibitemShut {NoStop}%
\end{thebibliography}%


\appendix

\section{4D Subspace for Third-Order Kronecker Graphs}

\begin{table*}
\caption{\label{table:subspace} For the third-order Kronecker graph $K_M \otimes K_M \otimes K_M$, the types of vertices that are nonadjacent to the marked vertex, how many such vertices there are, and the number of mutual neighbors each vertex has with the marked vertex.}
\begin{ruledtabular}
\begin{tabular}{lcc}
	Description of Nonadjacent Vertex & Number of Vertices & Number of Mutual Neighbors \\
	\colrule
	Same set, same subset, different position & $M-1$ & $(M-1)^2(M-2)$ \\
	Same set, different subset, same position & $M-1$ & $(M-1)^2(M-2)$ \\
	Same set, different subset, different position & $(M-1)^2$ & $(M-1)(M-2)^2$ \\
	Different set, same subset, same position & $M-1$ & $(M-1)^2(M-2)$ \\
	Different set, same subset, different position & $(M-1)^2$ & $(M-1)(M-2)^2$ \\
	Different set, different subset, same position & $(M-1)^2$ & $(M-1)(M-2)^2$ \\
\end{tabular}
\end{ruledtabular}
\end{table*}

Recall for the third-order Kronecker graph with complete initiator that there are four types of vertices. Let us begin by determining how many of each type of vertex there is using the $M = 4$ case in Fig.~\ref{fig:K4K4K4_v1_subspace} and then generalizing to arbitrary $M$.

The first type of vertex is the unique marked vertex, say vertex 1, so there is only one vertex of the first type. Next, we count the number of vertices adjacent to vertex 1, which are the vertices in a different set, different subset, and different position within the subset from the marked vertex. Vertex 1 has adjacent vertices in $(4-1) = 3$ sets, namely in $P_2$, $P_3$, and $P_4$. Within each of these sets, vertex 1 has adjacent vertices in $(4-1) = 3$ of the subsets, like $P_{2,2}$, $P_{2,3}$, and $P_{2,4}$. Within a subset, vertex 1 is adjacent to $(4-1) = 3$ vertices, such as vertices 22, 23, and 24. So vertex 1 has $(4-1)^3 = 27$ neighbors. Generalizing this, there are $(M-1)^3$ vertices adjacent to the marked vertex, and these are vertices of the second type.

The third and fourth types of vertices are nonadjacent to the marked vertex. Let us tabulate the vertices that are nonadjacent to the marked vertex, showing that they either share $(M-1)^2(M-2)$ mutual neighbors with the marked vertex or $(M-1)(M-2)^2$ mutual neighbors with the marked vertex.

Vertices that are in the same subset as the marked vertex are nonadjacent to the marked vertex. For example, in Fig.~\ref{fig:K4K4K4_v1_subspace}, if vertex 1 is marked, then vertices 2, 3, and 4 are in the same subset as vertex 1, and they are nonadjacent to vertex 1. In general, there are $(M-1)$ such vertices. Let us take vertex 2, for example. Vertices 1 and 2 share no common neighbors in $P_1$ since they are nonadjacent to all vertices in $P_1$. Within each of the $(4-1) = 3$ remaining sets $P_2$, $P_3$, and $P_4$, they share adjacent vertices in $(4-1) = 3$ subsets, such as $P_{2,2}$, $P_{2,3}$, and $P_{2,4}$ in $P_2$. Within each subset, there are $(4-2) = 2$ mutually adjacent vertices, such as vertices $23$ and $24$ in $P_{2,2}$. So, the total number of common neighbors is $(4-1)(4-1)(4-2) = 18$. Generalizing this, there are $(M-1)(M-1)(M-2) = (M-1)^2(M-2)$ mutual neighbors. This is summarized in the first body row of Table~\ref{table:subspace}.

A vertex that is nonadjacent to the marked vertex could be in the same set as the marked vertex, but a different subset, yet in the same position within its subset as the marked vertex is within its subset. In Fig.~\ref{fig:K4K4K4_v1_subspace}, these vertices would be vertices 5, 9, and 13. In general, there are $(M-1)$ such nonadjacent vertices. Let us take vertex 5, for example. Vertices 1 and 5 share no common neighbors in $P_1$, so $(4-1) = 3$ sets remain where they can have common neighbors. Within these sets, say $P_2$, there are no common neighbors in the top two subsets $P_{2,1}$ or $P_{2,2}$. This leaves $(4-2) = 2$ subsets $P_{2,3}$ and $P_{2,4}$ that contain mutual neighbors, and within each of these subsets, such as $P_{2,3}$, there are $(4-1) = 3$ mutual neighbors. So the total number of common neighbors is $(4-1)(4-2)(4-1) = 18$, as in the previous case, or $(M-1)(M-2)(M-1) = (M-1)^2(M-2)$ in general. This is summarized in the second body row of Table~\ref{table:subspace}.

Another vertex that is nonadjacent to the marked vertex is one in in the same set as the marked vertex, but a different subset, and in a different position. In Fig.~\ref{fig:K4K4K4_v1_subspace}, these are vertices 6, 7, 8, 10, 11, 12, 14, 15, and 16, and in general there are $(M-1)^2$ such vertices. Let us take vertex $6$, for example. Vertices 1 and 6 have mutual neighbors in the three remaining sets $P_2$, $P_3$, and $P_4$. In general, there are $(M-1)$ such sets. With a set, say $P_2$, there are $(4-2) = 2$ subsets $P_{2,3}$ and $P_{2,4}$ that contain mutual neighbors of vertices 1 and 6. In general, there are $(M-2)$ such subsets within each set. Finally, within each subset, say within $P_{2,3}$, vertices 27 and 28 are mutual neighbors of vertices 1 and 6, and in general, there are $(M-2)$ mutual neighbors in each subset. Altogether, there are $(M-1)(M-2)(M-2) = (M-1)(M-2)^2$ mutual neighbors. This is summarized in the third body row of Table~\ref{table:subspace}.

A vertex in a different set from the marked vertex, but in the same position, is also nonadjacent to the marked vertex. In Fig.~\ref{fig:K4K4K4_v1_subspace}, these vertices would be vertices 17, 33, and 49, and in general there are $(M-1)$ such vertices. How many mutual neighbors does one of these vertices have with the marked vertex? There are $(M-2)$ sets that contain mutual neighbors, each with $(M-1)$ subsets that contain mutual neighbors, each with $(M-1)$ mutually adjacent vertices, for a total of $(M-2)(M-1)(M-1) = (M-1)^2(M-2)$ mutual neighbors. This is summarized in the fourth body row of Table~\ref{table:subspace}.

One more kind of vertex that is nonadjacent to the marked vertex is one in a different set from the marked vertex, the same relative subset, but a different position within the subset. In Fig.~\ref{fig:K4K4K4_v1_subspace}, this corresponds to vertices 18, 19, 20, 34, 35, 36, 50, 51, and 52. In general, there are $(M-1)^2$ such vertices. Taking one of these vertices and the marked vertex 1, there are $(M-2)$ sets containing mutual neighbors, each with $(M-1)$ subsets containing mutual neighbors, each with $(M-2)$ mutual neighbors, for a total of $(M-1)(M-2)^2$ mutual neighbors. This is summarized in the fifth body row of Table~\ref{table:subspace}.

Finally, the last kind of vertex that is nonadjacent to the marked vertex lies in a different set, in a different relative subset, but the same position within the subset as the marked vertex. In Fig.~\ref{fig:K4K4K4_v1_subspace}, this corresponds to vertices 21, 25, 29, 37, 41, 45, 53, 57, and 61. In general, there are $(M-1)^2$ such vertices. Taking one of these vertices and the marked vertex 1, there are $(M-2)$ sets containing mutual neighbors, each with $(M-2)$ subsets containing mutual neighbors, each with $(M-1)$ mutual neighbors, for a total of $(M-1)(M-2)^2$ mutual neighbors. This is summarized in the last body row of Table~\ref{table:subspace}.

Summarizing these results, in total, there are $3(M-1)$ vertices that are nonadjacent to the marked vertex, each with $(M-1)^2(M-2)$ mutual neighbors with the marked vertex. Also, in total, there are $3(M-1)^2$ vertices that are nonadjacent to the marked vertex, each with $(M-1)(M-2)^2$ mutual neighbors with the marked vertex. This divides the vertices that are nonadjacent to the marked vertex into two Type 3 and Type 4 vertices.

Lastly, as a sanity check, we add the one marked vertex, the number of adjacent vertices, and the number of nonadjacent vertices of each type:
\[ 3(M-1) + 3(M-1)^2 + (M-1)^2 + 1 = M^3 = N. \]
We get the total number of vertices, as expected.

\end{document}